\documentclass[twocolumn,english,aps,pra,showpacs]{revtex4}
\usepackage[T1]{fontenc}
\usepackage[latin1]{inputenc}
\usepackage{amsmath}
\usepackage{graphicx}
\usepackage{amssymb}
\usepackage{esint}

\makeatletter

\def\k{\textbf{k}}

\makeatother

\usepackage{babel}

\begin{document}

\title{Finite-Momentum Dimer Bound State
in Spin-Orbit Coupled Fermi Gas}

\author{Lin Dong,$^{1}$ Lei Jiang,$^{2}$ Hui Hu,$^{3}$ Han Pu$^{1}$}

\affiliation{$^{1}$Department of Physics and Astronomy, and Rice Quantum Institute,
Rice University, Houston, Texas 77251, USA \\
 $^{2}$Joint Quantum Institute, University of Maryland and National
Institute of Standards and Technology, Gaithersburg, Maryland 20899,
USA \\
$^{3}$ARC Centre of Excellence for Quantum-Atom Optics, Centre for
Atom Optics and Ultrafast Spectroscopy, Swinburne University of Technology,
Melbourne 3122, Australia }

\date{\today}
\begin{abstract}
We investigate the two-body properties of a spin-1/2 Fermi gas subject to a spin-orbit coupling induced by laser fields. When an attractive $s$-wave interaction between unlike spins is present, the system may form a dimer bound state. Surprisingly, in the presence of a Zeeman field along the direction of the spin-orbit coupling, the bound state obtains finite center-of-mass mechanical momentum, whereas under the same condition but in the absence of the two-body interaction, the system has zero total momentum. This unusual result can be regarded as a consequence of the broken Galilean invariance by the spin-orbit coupling. Such a finite-momentum bound state will have profound effects on the many-body properties of the system.   
\end{abstract}

\pacs{05.30.Fk, 03.75.Hh, 03.75.Ss, 67.85.-d}

\maketitle

\section{Introduction}
With the experimental achievement of realizing the synthetic gauge field
in ultracold neutral atoms \cite{lin}, there has been an intensive
search for the emergence of exotic quantum phases \cite{2b,YuZhai, Iskin, Shenoy, Subasi, VictorGalitski,lei, Zhai2010,vj2011, GongZhang, HuiPu}.
So far, both Abelian and non-Abelian gauge fields have been realized
for quantum gases using the two-photon Raman process \cite{lin, Zhang, mit}. The realized non-Abelian gauge field leads to a particular spin-orbit coupling (SOC), which can be regarded as an equal-weight combination of Rashba and Dresselhaus SOC. Schemes of generating various other types of SOC have been proposed (e.g., \cite{Jacob2007, RMP, 3DSOC,zhou}) and are under active experimental investigation. Previous theoretical studies have shown that the interplay between SOC and effective Zeeman field (induced by the same laser fields and/or real magnetic fields) can lead to intriguing quantum states with nontrivial topological properties. However, the conditions under which such states can be realized have not been met in current experiments so far.

In this paper, we focus on the two-body properties of a degenerate spin-1/2
Fermi gas subject to spin-orbit coupling and effective Zeeman field. The bound states formed by two fermions subject to SOC have been studied rather intensively \cite{2b,YuZhai, Iskin, Shenoy, Subasi, VictorGalitski,lei}. However, in previous studies, the interplay between the SOC and the Zeeman field has not been thoroughly investigated. We show in the present work that, in the absence of two-body interaction, such a system exhibits zero total momentum. By contrast, in the presence of attractive $s$-wave interaction and with a proper combination of the SOC and the Zeeman field, the system may form a dimer bound state with {\em finite} center-of-mass mechanical momentum. At first sight, this result is very surprising since the $s$-wave interaction is manifestly momentum-conserving. A closer examination will reveal that this is a natural consequence of the broken Galilean invariance due to the presence of the SOC \cite{deviatedipole, ram, brokeGal}. In principle, this finite-momentum dimer state can be realized in the current experimental system with equal-weight combination of Rashba and Dresselhaus SOC. The momentum of the bound state can be measured through standard time-of-flight techniques.    

The paper is organized as follows: After introducing the general formalism of the two-body problem in Sec. II, we apply this formalism to experimentally realized equal-weight Rashba-Dresselhaus SOC in Sec. III. We briefly mention other types of SOC in Sec. IV, and finally conclude in Sec. V. 

\section{General Formalism}
We start by formulating the single-particle Hamiltonian for a noninteracting
homogeneous Fermi gas in three dimensions:  
\begin{equation}
\mathcal{H}_{0}=\frac{\hbar^2 {\bf k}^{2}}{2m}+\sum_{i=x,y,z}\left(v_{i}k_{i}+\Lambda_{i}\right)\sigma_{i} \,,
\label{eq:modelSingleH0}
\end{equation}
where we have defined SOC strength vector ${\bf v}=(v_{x},v_{y},v_{z})$,
and Zeeman field vector $\boldsymbol{\Lambda}=(\Lambda_x, \Lambda_y, \Lambda_z)$. $\boldsymbol{\sigma}=(\sigma_{x},\sigma_{y},\sigma_{z})$
are Pauli matrices acting on the atomic (pseudo-)spin degrees of freedom. This description is a general model valid for various coupling schemes. 
The single-particle spectrum can be straightforwardly obtained as 
$E_{\bf k}^{\pm}=\epsilon_\k \pm\sqrt{\sum_{i=x,y,z}(v_{i}k_{i}+\Lambda_{i})^{2}}$, with $\epsilon_\k = \hbar^2 \k^2/(2m)$. 
The superscripts `$\pm$' define the two helicity bases which are related to the original spin basis by the transformation 
\begin{equation}
\left[\begin{array}{c}
|{\bf k}+\rangle\\
|{\bf k}-\rangle\end{array}\right]=\left[\begin{array}{cc}
\cos\theta_{\bf k} & \sin\theta_{\bf k} \, e^{i\phi_{\bf k}}\\
-\sin\theta_{\bf k} \, e^{-i\phi_{\bf k}} & \cos\theta_{\bf k}\end{array}\right]\left[\begin{array}{c}
|{\bf k}\uparrow\rangle\\
|{\bf k}\downarrow\rangle\end{array}\right] \,,
\end{equation}
where
\begin{eqnarray*}
\cos^2 \theta_{\bf k} &= & \frac{1}{2}\left(1+\frac{v_{z}k_{z}+\Lambda_z}{\sqrt{\sum_{i=x,y,z}(v_{i}k_{i}/m+\Lambda_{i})^{2}}}\right) \,, \\
\tan\phi_{\bf k} &=& -\frac{v_{y}k_{y}+\Lambda_y}{v_{x}k_{x}+\Lambda_x}\,.
\end{eqnarray*}


Next we consider the
attractive $s$-wave contact interaction between unlike spins which, in terms of the creation and annihilation operators for the original spin states, are represented by
\begin{equation}
\mathcal{H}_{\rm int}=\frac{g}{V}\sum_{{\bf k}{\bf k^{\prime}}{\bf q}}c_{{\bf q/2+k}\uparrow}^{\dag}c_{{\bf q/2-k}\downarrow}^{\dag}c_{{\bf q/2-k^{\prime}}\downarrow}c_{{\bf q/2+k^{\prime}}\uparrow} \,,\label{int}
\end{equation}
where $g$ is the bare coupling strength to be renormalized using the $s$-wave scattering length $a_s$.
A general two-body wave function describing a dimer state with a definite center-mass momentum ${\bf q}$ can be written as \cite{YuZhai}
\begin{equation}
|\Psi\rangle_{{\bf q}}={\sum_{{\bf k}, \sigma, \sigma'}}\,\psi_{\sigma \sigma'}({\bf k}) \,c_{\frac{{\bf q}}{2}+{\bf k}\sigma}^{\dagger}c_{\frac{{\bf q}}{2}-{\bf k}\sigma'}^{\dagger}\,|0\rangle \,. \label{2body}
\end{equation}
Inserting this wave function into the Schr\"{o}dinger
equation 
\begin{equation}
(\mathcal{H}_{0}+\mathcal{H}_{{\rm int}}\big)|\Psi\rangle_{{\bf q}}=E_{{\bf q}}|\Psi\rangle_{{\bf q}}  \,, \label{Schrodingerequation}
\end{equation} and after some lengthy algebra, we arrive at four
coupled algebraic equations for the coefficients $\psi_{\sigma \sigma'}({\bf k})$. Self-consistency requires that the energy of the dimer state $E_{\bf q}$, satisfies the following equation:
\begin{widetext}
\begin{equation}
\frac{m}{4\pi\hbar^{2}a_{s}}=\frac{1}{V}\sum_{\bf k}\left[\left({\mathcal{E}_{{\bf k,q}}-\frac{4\mathcal{E}_{{\bf k,q}}^{2}({\bf v}\cdot {\bf k})^{2}-4\left[\sum_{i=x,y,z} v_{i}k_{i}(v_{i}q_{i}+2\Lambda_{i}) \right]^{2}}{\mathcal{E}_{{\bf k,q}}\left(\mathcal{E}_{{\bf k,q}}^{2}-\sum_{i=x,y,z}(v_{i}q_{i}+2\Lambda_{i})^{2}\right)}}\right)^{-1}+\frac{1}{2\epsilon_{\bf k}}\right]\,,\label{Cooper}
\end{equation}
where $\mathcal{E}_{{\bf k,q}} \equiv E_{{\bf q}}-\epsilon_{\frac{{\bf q}}{2}+{\bf k}}-\epsilon_{\frac{{\bf q}}{2}-{\bf k}}$, and the interaction has been regularized as $1/g=m/(4\pi \hbar^2 a_{{s}})-1/V\sum_{{\bf k}}1/(2\epsilon_{{\bf k}})$. The explicit forms of $\psi_{\sigma \sigma'}(\k)$ can also be found, which we list in Appendix~\ref{secA}.

\end{widetext}
\section{Equal-weight Rashba-Dresselhaus SOC}
We now apply this general formalism to the experimentally realized system whose single-particle Hamiltonian takes the following form:
\begin{eqnarray}
H_{0}  &=&   \sum_\sigma \int d{\bf r} \Psi_{\sigma}^{\dagger}({\bf r})\left(\frac{\hbar^{2}{\bf k}^{2}}{2m}+\alpha\delta\right)\Psi_{\sigma}({\bf r}) \nonumber\\
&+& \int d{\bf r}  \left(\frac{\Omega}{2}e^{2i\hbar k_{r}z}  \Psi_{\uparrow}^{\dagger}({\bf r}) \Psi_{\downarrow}({\bf r})+h.c.\right) \,.
\end{eqnarray}
Here $\Psi_{\sigma}$ and $\Psi_{\sigma}^\dag$ are field operators for hyperfine spin states $\sigma$, and $\alpha=\pm1$ for $\sigma=\uparrow, \downarrow$, respectively. $\Omega$ is the two-photon Raman coupling strength and $k_r$ the recoil momentum of the Raman beams, which is taken to be along the $z$ axis. Finally, $\delta$ represents the two-photon detuning. To get rid of the exponential terms, we introduce a local gauge transformation:
\begin{equation}
\tilde{\psi}_\uparrow ({\bf r})= e^{-i k_r z} \Psi_\uparrow({\bf r})\,,\;\; \tilde{\psi}_\downarrow ({\bf r}) = e^{i k_r z} \Psi_\downarrow ({\bf r})\,. \label{gauge}
\end{equation}
Defining the spinor $\tilde{\boldsymbol \psi} = (\tilde{\psi}_\uparrow ,\;\tilde{\psi}_\downarrow)^T$, we can recast $H_0$ as 
\begin{eqnarray}
H_0 &=& \int d{\bf r} \, \tilde{\boldsymbol \psi}^\dag  ({\bf r})\, {\cal H}_0 \,\tilde{\boldsymbol \psi}  ({\bf r}) \,, \\
{\cal H}_0 &=& \frac{\hbar^2 {\bf k}^2}{2m} + \frac{\hbar^2 k_r}{m} k_z \sigma_z + \frac{\Omega}{2} \sigma_x + \delta \sigma_z \,,\label{h0}
\end{eqnarray}
where we have neglected a constant energy shift $E_r={\hbar^2 k_r^2}/{(2m)} $ (the recoil energy) in ${\cal H}_0$. One can immediately see that ${\cal H}_0$ above takes the form of Eq.~(\ref{eq:modelSingleH0}), with ${\bf v}=(0,\;0,\; \hbar^2 k_r/m)$ and ${\boldsymbol \Lambda} = (\Omega/2, \;0,\; \delta)$. The two-body interaction Hamiltonian takes the form of Eq.~(\ref{int}) after we define the momentum space operators via $\tilde{\psi}_\sigma = \sum_{\bf k} e^{i{\bf k} \cdot {\bf r}} c_{{\bf k} \sigma} /\sqrt{V}$, and is obviously invariant under the transformation (\ref{gauge}).

\subsection{Single-Particle Ground State}
Before we discuss the two-body properties of the interacting system, let us briefly examine the noninteracting limit. The single-particle ground state occurs in the lower helicity branch and is given by $|{\bf k}_{\rm min} - \rangle$ at momentum ${\bf k}_{\rm min} = (0,\,0,\, k_0)$ with energy 
\begin{equation}
E_{\rm min} = \frac{\hbar^2 k_0^2}{2m} -\sqrt{h^2 +(\lambda k_0+\delta)^2} \,, \label{Emin}
\end{equation}
where for notational simplicity we have defined $h \equiv \Omega/2$ and $\lambda \equiv \hbar^2 k_r/m$. From ${\partial E_{\rm min}}/{\partial k_0}  =0 $, we obtain 
\begin{equation}
k_{0}=\frac{\lambda k_{0}+\delta}{\sqrt{h^{2}+(\lambda k_{0}+\delta)^{2}}} \, k_r \,.\label{eq:grstk}\end{equation}
It is important to remember that ${\bf k}_{\rm min}$ is not the {\em mechanical} momentum of the particle.
 Measured in the laboratory frame, the mechanical momentum of the particle prepared in this ground state can be calculated as 
\begin{equation}
{\bf K}_0 = ({\bf k}_{\rm min} + k_r \hat{z}) \sin^2 \theta_{{\bf k}_{\rm min}} +  ({\bf k}_{\rm min} - k_r \hat{z}) \cos^2 \theta_{{\bf k}_{\rm min}} \,,\label{po}
\end{equation}
where the two terms proportional to $k_r$ are included to ``undo'' the local gauge transformation performed in Eq.~(\ref{gauge}). After some straightforward algebra, we can show that ${\bf K}_0=0$, i.e., particle in the ground state has exactly zero mechanical momentum in the laboratory frame, even though its canonical momentum $k_0$ depends explicitly on the strengths of the SOC $\lambda$, and the Zeeman fields $h$ and $\delta$. 
\begin{figure}[htp]
\includegraphics[clip,width=.48\textwidth]{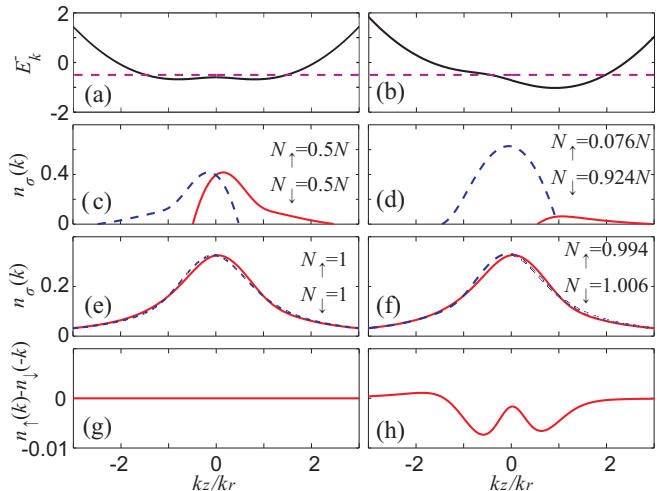} 

\caption{(Color online) (a, b) Single-particle dispersion in the lower helicity branch. The horizontal dashed lines represent the chemical potential of the Fermi sea. Energy is in units of $E_r$. (c, d) Momentum distribution along the $z$ axis for the noninteracting Fermi gas is represented by (a) and (b), respectively. The relative population of the two spin components are listed in the plot, blue dashed line for spin-down atoms, and red thick line for spin-up atoms. (e, f) Momentum distribution along the $z$ axis of the dimer bound state. (g, h) Difference between $n_\uparrow(k_z)$ and $n_\downarrow(-k_z)$ for the cases shown in (e) and (f), respectively. For the left column [(a), (c), (e), and (g)] we have $\delta=0$ and $h=0.6E_r$; for the right column [(b), (d), (f), and (h)] we have $\delta=0.4E_r$ and $h=0.6E_r$. For the interacting case [(e)--(h)], the interaction strength is given by $1/(k_r a_s)=1$.}

\label{fig1} 
\end{figure}

\subsection{Non-Interacting Fermi Sea}
Next we consider a filled Fermi sea at zero temperature. Instead of taking ${\bf k}={\bf k}_{\rm min}$ as in Eq.~(\ref{po}), we need to sum over all the momentum under the Fermi surface. Here again one can show analytically that the total momentum of the Fermi gas is zero in the laboratory frame (see Appendix~\ref{secC}). The single-particle energy dispersion and momentum distributions of two examples of the noninteracting Fermi sea (one for $\delta=0$ and the other $\delta \neq 0$) are illustrated in Figs.~\ref{fig1}(a)--1(d). The momentum distribution plotted in the figure can be readily measured using a spin-resolved time-of-flight technique that is a quite standard tool in cold atom experiments.

\subsection{Two-Body Bound State of Interacting System}
\begin{figure}[htp]
\includegraphics[clip,width=.4\textwidth]{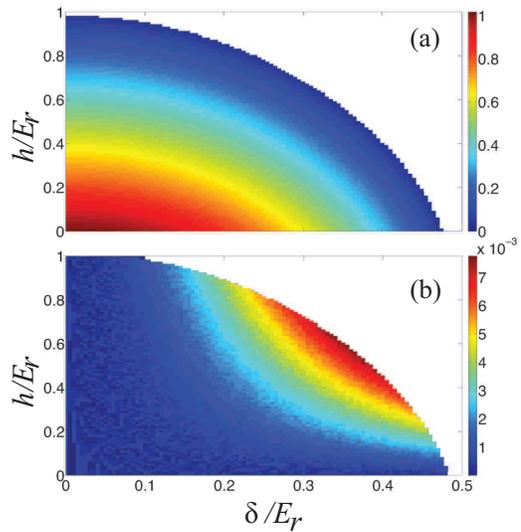} 

\caption{(Color online) Binding energy $\epsilon_b$ and the magnitude of the bound-state momentum $q_0$ as functions of Zeeman field strengths $\delta$ and $h$. The coloring in (a) represents $\epsilon_b/E_r$, and that in (b) represents $q_0/k_r$. The white region is where no bound states can be found. 
The scattering length is given by $1/(k_r a_s)=1$.}

\label{fig2} 
\end{figure}

Now we are in the position to discuss the two-body results when interaction is included. Following the general formalism, for a given set of parameters $h$, $\delta$, and $a_s$, we can obtain numerically the eigenenergy of the dimer state $E_{\bf q}$ as a function of ${\bf q}$. The momentum ${\bf q}_0$ at which $E_{\bf q}$ reaches the minimum labels the ground dimer state. The binding energy is defined as \begin{equation}
\epsilon_b = 2E_{\rm min} - E_{{\bf q}_0}\,. \label{eb}
\end{equation}
Only when $\epsilon_b >0$, can we consider the dimer as a truly two-body bound state. Otherwise, its energy lies in the single-particle continuum. Figure \ref{fig2}(a) shows $\epsilon_b$ decreases with increasing $h$ and $\delta$. Beyond a critical boundary value, binding energy becomes negative and no stable bound state can be found. For this system, a two-body bound state only occurs on the BEC side of a Feshbach resonance with $a_s >0$ \cite{Shenoy}. More importantly, ${\bf q}_0 = q_0 \hat{z}$ will be {\em nonzero} and along the $z$ axis as long as both $h$ and $\delta$ are finite. Figure \ref{fig2}(b) displays $q_0$ as functions of $h$ and $\delta$. We note that ${\bf q}_0$ is an even function of $h$ and an odd function of $\delta$. In Appendix~\ref{proof}, we prove that for a given nonzero $h$, $q_0$ deviates from zero for arbitrarily small $\delta$.

%
%
%

For the two-body wave function given in Eq.~(\ref{2body}), the momentum distribution of the hyperfine spin states in the laboratory frame is given by
\begin{eqnarray}
n_\uparrow ({\bf k}+ k_r \hat{z})\!\! &=& \!\! \langle c^\dag_{{\bf k} \uparrow}c_{{\bf k} \uparrow} \rangle \nonumber \\  &=& |\psi_{\uparrow \downarrow}({\bf k}-{\bf q}/2)|^2 + |\psi_{\uparrow \uparrow}({\bf k}-{\bf q}/2)|^2 \,, \\
n_\downarrow ({\bf k}- k_r \hat{z}) \!\! &=& \!\! \langle c^\dag_{{\bf k} \downarrow}c_{{\bf k} \downarrow} \rangle \nonumber \\ &=& |\psi_{\downarrow \uparrow}({\bf k}-{\bf q}/2)|^2 + |\psi_{\downarrow \downarrow}({\bf k}-{\bf q}/2)|^2 \,.
\end{eqnarray}
Both of these distribution functions will be symmetric along the $x$ and $y$ axis, which yields the average momentum along these two axes $P_{x,y}^\sigma= \int d{\bf k} \, k_{x,y}n_\sigma({\bf k}) =0$. By contrast, $n_\sigma ({\bf k})$ will be in general asymmetric along the $z$-axis which results in finite values of $P_z^\sigma = \int d{\bf k} \, k_{z}n_\sigma({\bf k})$. The total momentum 
can be shown as 
\begin{eqnarray}
{\bf K}_{\rm{lab}} &=& \sum_{\bf k} \left[ {\bf k} \,(n_\uparrow({\bf k}) + n_\downarrow({\bf k})) \right] \nonumber \\
&=& {\bf q} + (N_\uparrow -N_\downarrow)k_r\hat{z} \,,\label{pz}
\end{eqnarray} 
where $N_\sigma = \int d{\bf k} \,n_\sigma({\bf k})$ is the population in spin-$\sigma$ and satisfy the obvious constraint $N_\uparrow+N_\downarrow =2$. In Fig.~\ref{fig1}(e) and (f), we illustrate how $n_\sigma (k_z) = \int dk_x \int dk_y \, n_\sigma ({\bf k})$ changes without and with detuning $\delta$. To see it more clearly, we plot the difference between $n_\uparrow(k_z)$ and $n_\downarrow(-k_z)$ in Fig.~\ref{fig1}(g) and (h). Notice that it is the SOC that breaks spatial reflectional symmetry such that $n_\sigma({\bf k})\neq n_\sigma({\bf -k})$ with $\sigma=\uparrow,\downarrow$. However, for $\delta=0$, one still has the symmetry $n_\uparrow ({\bf k})= n_\downarrow (-{\bf k})$ as measured in the experiment of Ref.~\cite{Zhang}; with finite $\delta$, this symmetry between $n_\uparrow(\bf k)$ and $n_\downarrow(-{\bf k})$ is further broken. The momentum distribution for the interacting system can be obtained using the same time-of-flight technique as we mentioned earlier. However, before the atoms are released, one needs to turn off the interaction via, e.g., Feshbach resonace, so that the time-of-flight images provide momentum information for the initial bound state.  

Note that for either $h=0$ or $\delta=0$, we obtain $q_0=0$ and $N_\uparrow =N_\downarrow=1$. In either of these cases, both $P_z^\uparrow$ and $P_z^\downarrow$ will be finite, but they have equal magnitude and opposite sign and hence the total momentum of the dimer $P_z =0$. When both $h$ and $\delta$ are non-zero, we obtain finite $q_0$ and moreover $N_\uparrow \neq N_\downarrow$. Our numerical calculation shows that, under such circumstances, $P_z \approx 2q_0$.

\begin{figure}[htp]
\includegraphics[width=.48\textwidth]{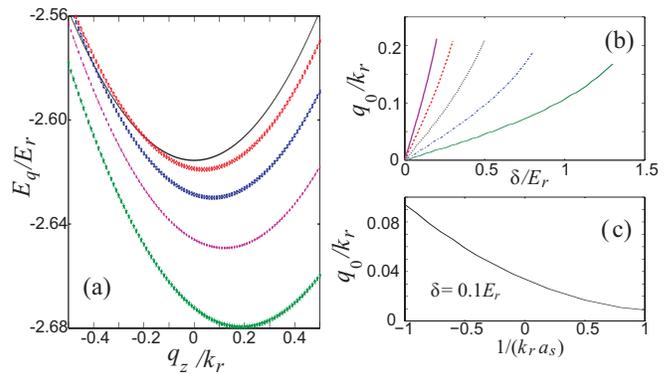} 

\caption{ (Color online) (a) For spherical SOC, bound-state energy $E_q$ as a function of $q$ (momentum along the $z$ axis) for different values of Zeeman field strength, from top to bottom $\delta/E_r=0,0.2,0.4,0.6,0.8 $ and $1/(k_r a_s)=1/2$ for all curves. (b) Ground-state momentum $q_0$ as a function of  $\delta$ at different values of scattering length. From left to right, the curves correspond to  $1/(k_r a_s)=-1$, $-1/2$, 0, 1/2, and 1, respectively. (c) $q_0$ as a function of $1/(k_r a_s)$ at $\delta/E_r=0.1$. }

\label{fig3} 
\end{figure} 

This situation, combined with the results for the non-interacting system, leads to the following peculiar phenomenon: In the absence of two-body interaction, the system possesses zero total momentum; when the interaction is turned on, the system picks up a finite total momentum even though the interaction Hamiltonian (\ref{int}) seems to be momentum-conserving. This peculiarity is indeed a salient feature of the SOC. Since the linear momentum of the atom is intimately coupled to its (pseudo-) spin, as the interaction induces redistribution of atomic population (i.e., changes the value of $\sigma_z$) as shown in Fig.~\ref{fig1}(d) and (f), it also modifies the total momentum of the system. This phenomenon can be regarded as a manifestation of the broken Galilean invariance, which is also responsible for other unusual behaviors such as the deviation of dipole oscillation frequency in a harmonically trapped system \cite{deviatedipole, ram}, and the ambiguity in defining Landau critical velocity in spin-orbit coupled condensates \cite{brokeGal}.

\section{Other Types of SOC}
The momentum of each spin species can be readily measured in experiment using the time-of-flight technique. This has actually been performed in a non-interacting SOC Fermi gas \cite{Zhang}. However, our calculation shows that $q_0$ is only on the order  1\% of recoil momentum for the equal Rashba-Dresselhaus SOC (see Fig.~\ref{fig2}) which is the only type of SOC realized so far. The magnitude of $q_0$, however, can be greatly enhanced under other SOC scheme. As an example, we consider the spherical SOC coupling scheme that is recently proposed \cite{3DSOC,zhou}. For this case, the SOC strength $v_x=v_y=v_z=v$. We take the spin-orbit coupling strength $v=\hbar^2 k_r/m$ to be the same as in the previous case. Due to the isotropic nature of the SOC term, the direction of the Zeeman field is irrelevant. We choose it to be along the $z$-axis, i.e., ${\boldsymbol \Lambda}=(0,\,0,\,\delta)$, which, as we shall show, leads to a dimer state with finite moment $q_0 \hat{z}$ along the $z$-axis.  Fig.~\ref{fig3}(a) demonstrates that, with increasing $\delta$, the minimum of of the bound state energy $E_q$ deviates away from zero to some finite value. Compared to the former equal Rashba-Dresselhaus SOC case, an essential difference here is that a two-body bound state can be realized even when $a_s<0$ \cite{vj2011}.  Fig.~\ref{fig3}(b) and (c) show how $q_0$ varies as functions of $\delta$ and $a_s$. As one can see, as long as $\delta$ is non-zero, the dimer ground state possesses finite momentum. Furthermore, The magnitude of $q_0$ can reach as high as $0.2k_r$. Such a large value should be easily detected in experiment.

Finally, for the sake of completeness, we comment on the case of Rashba SOC with ${\bf v}=(v_x,\,v_y,\,0)$ and $v_x=v_y=\hbar^2 k_r/m$. In this case, when an in-plane Zeeman field (i.e., with a component in the $x$-$y$ plane) is present, the resulting dimer bound state will again have finite momentum. A plot similar to Fig.~\ref{fig3} can be obtained. Our calculation shows that the maximum $q_0$ can be reached is about $0.05k_r$. 

\section{Conclusion}
In summary, we have studied the two-body properties of a degenerate spin-1/2 Fermi gas subjected to spin-orbit coupling and effective Zeeman field. We show that, two fermions, via $s$-wave scattering, may form a dimer bound state with $\emph {finite}$ center-of-mass mechanical momentum, under proper configuration of SO coupling and Zeeman field. We attribute this peculiar phenomenon to the broken Galilean invariance induced by the SO coupling and special role played by the Zeeman field. We have directly solved the two-body Schr\"{o}dinger equation in the general formalism, and considered the momentum distribution in the laboratory frame for the experimentally realized system, where finite but relatively small bound state momentum $q_0$ is found.  Finally, we demonstrate that the recently proposed system with spherical SOC \cite{3DSOC,zhou} can result in dimer bound state with up to $20\%$ of the recoil momentum. Note that, for bosons, previous studies have shown in the presence of SOC, ground state is either plane wave phase or standing wave phase \cite{jason, zhaihui, yip}. However, in contrast to our work, such states possess exactly \emph{zero} mechanical momentum. In a many-body setting, our two-body calculation is relevant in the limit where most of the atoms form tightly bound pairs. When this limit is not reached, we still expect that the interplay between SOC and Zeeman field may lead to an exotic superfluid state. For instance, as a direct analog of the finite-momentum dimer, the many-body system may support finite-momentum Cooper pairs \cite{guo,vjnote,wei} reminiscent of the Fulde-Ferrell state studied in the context of Fermi gases with spin imbalance. A systematic study of such many-body properties will be conducted in the future. 

\textit{Acknowledgment} ---H.P. is supported by the NSF, the Welch Foundation (Grant No. C-1669), and DARPA. H.H. is supported by the ARC Discovery Projects (Grant No. DP0984522) and the National Basic Research Program of China (NFRP-China, Grant No. 2011CB921502). We would like to thank Vijay Shenoy, Biao Wu, Wei Yi, and Xiangfa Zhou for useful discussions, and Vijay Shenoy for sharing his note before publication \cite{vjnote}.


\appendix

\begin{widetext}
\section{RELEVANT EQUATIONS FOR GENERAL FORMALISM}\label{secA}
 In this Appendix, we derive the relevant equations for the two-body problem under the general SOC.
We start from the Schr\"{o}dinger equation, i.e., Eq.~(\ref{Schrodingerequation}) in the main text. Using the form of the Hamiltonian given in Eqs.~(\ref{eq:modelSingleH0}) and (\ref{int}), and that for the state vector $|\Psi \rangle_{\bf q}$ in Eq.~(\ref{2body}),
we can obtain four coupled equations for the coefficients $\psi_{\sigma \sigma'}$ that characterize the state $|\Psi \rangle_{\bf q}$. 
Let us introduce the singlet wave function $\psi_{s}(\k)=\frac{1}{\sqrt{2}}[\psi_{\uparrow\downarrow}(\k)-\psi_{\downarrow\uparrow}(\k)]$
and the triplet wavefunctions $\psi_{t}(\k)=\frac{1}{\sqrt{2}}[\psi_{\uparrow\downarrow}(\k)+\psi_{\downarrow\uparrow}(\k)]$,
$\psi_{u}=\frac{1}{\sqrt{2}}(\psi_{\uparrow\uparrow}(\k)+\psi_{\downarrow\downarrow}(\k))$,
$\psi_{v}=\frac{1}{\sqrt{2}}(\psi_{\uparrow\uparrow}(\k)-\psi_{\downarrow\downarrow}(\k))$. The four equations for $\psi_{\sigma \sigma'}$ can be recast into the following form:
\begin{eqnarray}
\mathcal{E}_{{\bf k,q}}\, \psi_{s}  &=& \frac{2g}{V}{\sum_{{\bf k}'}}' \psi_{s}({\bf k}')+2v_{z}k_{z}\psi_{t}-2v_{x}k_{x}\psi_{v}-2v_{y}ik_{y}\psi_{u} \,,\label{eq:psis}\\
{\bf M} \left( \begin{array}{c} \psi_t \\ \psi_u \\ \psi_v \end{array} \right) & =& \left( \begin{array}{c} 2 v_z k_z \\ 2i v_y k_y \\ -2v_x k_x \end{array} \right) \,\psi_s \,,
\end{eqnarray}
 where $\mathcal{E}_{{\bf k,q}}=E_{{\bf q}}-\epsilon_{\frac{{\bf q}}{2}+{\bf k}}-\epsilon_{\frac{{\bf q}}{2}-{\bf k}}$, and $\sum_{\bf k}'$ denotes summation over positive momentum
$k_{z}$, and the matrix {\bf M} is given by 
\[ {\bf M} = \left(\begin{array}{ccc}
\mathcal{E}_{{\bf k,q}} & -(v_{x}q_{x}+2\Lambda_x) & -i(v_{y}q_{y}+2\Lambda_y)\\
-(v_{x}q_{x}+2\Lambda_x) & \mathcal{E}_{{\bf k,q}} & -(v_{z}q_{z}+2\Lambda_z)\\
i(v_{y}q_{y}+2\Lambda_y) & -(v_{z}q_{z}+2\Lambda_z) & \mathcal{E}_{{\bf k,q}}\end{array}\right)\,.\]
Denoting \begin{eqnarray*} M &\equiv&
{\rm det}({\bf M})
=\mathcal{E}_{{\bf k,q}}\left[\mathcal{E}_{{\bf k,q}}^{2}-\sum_{i=1}^3(v_{i}q_{i}+2\Lambda_i)^{2}\right]\,,
\end{eqnarray*} and using Cramer's rule in linear algebra, we can express triplet components in terms of $\psi_s(\k)$ as
\begin{eqnarray*}
\psi_{t}(\k) & = & \frac{2\psi_{s}(\k)}{M} \left|\begin{array}{ccc}
v_{z}k_{z} & -(v_{x}q_{x}+2\Lambda_x) & -i(v_{y}q_{y}+2\Lambda_y)\\
iv_{y}k_{y} & \mathcal{E}_{{\bf k,q}} & -(v_{z}q_{z}+2\Lambda_z)\\
-v_{x}k_{x} & -(v_{z}q_{z}+2\Lambda_z) & \mathcal{E}_{{\bf k,q}}\end{array}\right| \,,\label{psit}\\
\psi_{u}(\k) & = & \frac{2\psi_{s}(\k)}{M}\left|\begin{array}{ccc}
\mathcal{E}_{{\bf k,q}} & v_{z}k_{z} & -i(v_{y}q_{y}+2\Lambda_y)\\
-(v_{x}q_{x}+2\Lambda_x) & iv_{y}k_{y} & -(v_{z}q_{z}+2\Lambda_z)\\
i(v_{y}q_{y}+2\Lambda_y) & -v_{x}k_{x} & \mathcal{E}_{{\bf k,q}}\end{array}\right| \,,\label{psiu}\\
\psi_{v}(\k) & = & \frac{2\psi_s(\k)}{M} \left|\begin{array}{ccc}
\mathcal{E}_{{\bf k,q}} & -(v_{x}q_{x}+2\Lambda_x) & v_{z}k_{z}\\
-(v_{x}q_{x}+2\Lambda_x) & \mathcal{E}_{{\bf k,q}} & iv_{y}k_{y}\\
i(v_{y}q_{y}+2\Lambda_y) & -(v_{z}q_{z}+2\Lambda_z) & -v_{x}k_{x}\end{array}\right|\,.
\label{psiv}
\end{eqnarray*} 
After inserting expressions of $\psi_{t}(\k)$, $\psi_{u}(\k)$, and $\psi_{v}(\k)$
into Eq.~(\ref{eq:psis}), and integrating both sides over momentum 
and dividing by the constant $\frac{g}{V}\sum_{\k}\psi_{s}(\k)$,
we finally reach Eq.~(\ref{Cooper}) in the main text.
The un-normalized singlet wave function is determined from Eq.~(\ref{eq:psis}) as 
\begin{equation}
\psi_{s}(\k)=\left[\mathcal{E}_{{\bf k,q}}-\frac{4\mathcal{E}_{{\bf k,q}}^{2}({\bf v} \cdot {\bf k})^{2}-4\left(\sum_{i=x,y,z}v_{i}k_{i}(v_{i}q_{i}+2\Lambda_{i})\right)^{2}}{\mathcal{E}_{{\bf k,q}}\left(\mathcal{E}_{{\bf k,q}}^{2}-\sum_{i=x,y,z}(v_{i}q_{i}+2\Lambda_{i})^{2}\right)}\right]^{-1}\,,
\end{equation}
and the triplet wave functions are obtained accordingly. 

\section{ TOTAL MOMENTUM OF NONINTERACTING FERMI GAS AT ZERO TEMPERATURE}\label{secC}
Now consider a noninteracting Fermi sea described by Hamiltonian (8) in the main text. 
For a given chemical potential $\mu$, the Fermi
surface is defined by the following equation (we take $\hbar=m=1$): \begin{equation}
\frac{k_{z}^{2}+k_{\perp}^{2}}{2}-\sqrt{h^{2}+(\lambda k_{z}+\delta)^{2}}=\mu \,,\label{mu}
\end{equation}
where we have assumed that $\mu$ lies below the bottom of the upper helicity branch
and the Fermi surface is simply connected, as illustrated in Figs. 1(a) and 1(b) of the main text. The total mechanical momentum in the laboratory frame is obtained by integrating
over the three-dimensional volume of Fermi sea: 
\begin{eqnarray*}
K_{\rm lab}^{z} & = & \sum_{\k} \left[|\sin\theta_{\k}|^{2}(k_{z}+k_{r})+|\cos\theta_{\k}|^{2}(k_{z}-k_{r}) \right] \\
 & = & \frac{V}{(2\pi)^{3}}\int_{k_{1}}^{k_{2}}dk_{z}\left[k_{z}-\frac{(\lambda k_{z}+\delta)k_{r}}{\sqrt{h^{2}+(\lambda k_{z}+\delta)^{2}}}\right]
\times \left(\mu-\frac{1}{2}k_{z}^{2}+\sqrt{h^{2}+(\lambda k_{z}+\delta)^{2}}\right)\\
 & = & \frac{V}{8(2\pi)^{3}}\left(k_{2}^{4}-k_{1}^{4}+8\delta\lambda(k_{1}-k_{2})-4(\mu+\lambda^{2})(k_{2}^{2}-k_{1}^{2})\right)\\
 & = & 0 \,,\end{eqnarray*}
where $k_{1,2}$ are the intersections of the Fermi level with the lower helicity branch on the $z$ axis and are simply the roots of Eq.~(\ref{mu}) after taking $k_\perp =0$.
Similarly, \begin{eqnarray*}
K_{\rm lab}^{x} & = & \sum_{\k} \left[ |\sin\theta_{\k}|^{2}k_{x}+|\cos\theta_{\k}|^{2}k_{x} \right] =0 \,,\\
K_{\rm lab}^{y} & = & \sum_{\k} \left[|\sin\theta_{\k}|^{2}k_{y}+|\cos\theta_{\k}|^{2}k_{y} \right] =0 \,.
\end{eqnarray*}

\section{PROOF FOR FINITE-MOMENTUM DIMER GROUND STATE}
\label{proof}
For the equal-weight Rashba-Dresselhaus SOC, by taking ${\bf v}=(0,\;0,\; \lambda)$ and ${\boldsymbol \Lambda} = (h, \;0,\; \delta)$, and considering the possibility of a bound state with momentum ${\bf q} = q\hat{z}$, the bound-state energy $E_{ q}$ satisfies
\begin{equation}
\frac{m}{4\pi\hbar^{2}a_{{\rm s}}}=\frac{1}{V}\sum_{{\bf k}}\left\{ \left[{\mathcal{E}_{{\bf k,q}}-\frac{4\lambda^2 k^2_{z}}{\mathcal{E}_{{\bf k,q}}} \, \frac{\mathcal{E}_{{\bf k,q}}^{2}-(\lambda q+2\delta)^{2}}{\mathcal{E}_{{\bf k,q}}^{2}-4h^{2}-(\lambda q+2\delta)^{2}}} \right]^{-1}+\frac{1}{2\epsilon_{{\bf k}}} \right\} \,, \label{eq}
\end{equation}
where $\mathcal{E}_{{\bf k,q}}=E_{{ q}}-\epsilon_{\frac{{\bf q}}{2}+{\bf k}}-\epsilon_{\frac{{\bf q}}{2}-{\bf k}}=E_{q}-(\frac{q^{2}}{4}+k_{\perp}^{2}+k_{z}^{2})$. As a first step, we show that the ground state occurs at $q=0$ when $\delta=0$. To this end, we need to prove that $dE_q/dq|_{q=0}=0$. To show this, we take the derivative with respect to $q$ on both sides of Eq.~(\ref{eq}) and take $q=0$, which yields
\begin{eqnarray}
0 &=& \left. \frac{dE_q}{dq} \right|_{q=0} \,\sum_{\bf k} \,{\cal A}({\bf k}) \,, \label{c2}\\ {\cal A}({\bf k}) &=& \frac{[(k^2-E_0)^2-4h^2)]^2+4\lambda^2 k_z^2[(k^2-E_0)^2+4h^2]}{(k^2-E_0)^2 [(k^2-E_0)^2-4h^2-4\lambda^2k_z^2]^2} \,.\end{eqnarray} 
Since the momentum integral in Eq.~(\ref{c2}) is finite as the integrand ${\cal A}({\bf k})$ is non-negative, we must have $dE_q/dq|_{q=0}=0$, indicating that the ground state indeed has zero momentum. 

Next, we turn to a finite but small $\delta$. We perform a similar calculation as above and expand all terms to first order in $\delta$. This leads to
\begin{eqnarray}
0&=& \left. \frac{dE_q}{dq} \right|_{q=0} \,\sum_{\bf k} \,{\cal A}({\bf k}) + \delta \,\sum_{\bf k}\,{\cal B}({\bf k}) \,,\label{c4} \\
{\cal B}({\bf k}) &=& \frac{64\lambda^3 h^2 k_z^2}{(k^2-E_0)^3 [(k^2-E_0)^2-4h^2-4\lambda^2k_z^2]^2} \,.
\end{eqnarray}
For a bound state, we have $E_0<0$; hence ${\cal B}({\bf k})$ is also non-negative. Consequently, we conclude from Eq.~(\ref{c4}) that $dE_q/dq|_{q=0} \propto \delta$. Therefore we have proved that for any finite $\delta$, the ground dimer state cannot have zero momentum.  

\end{widetext}


\begin{thebibliography}{4}

\bibitem{lin} Y.-J. Lin, R. L. Compton, A. R. Perry, W. D. Phillips, J. V. Porto, and I. B. Spielman, Phys. Rev. Lett. {\bf 102}, 130401 (2009); Y.-J. Lin, R. L. Compton, K. Jimenez-Garcia, J. V. Porto, and I. B. Spielman, Nature (London), {\bf 426}, 628 (2009); Y.-J. Lin, K. Jimenez-Garcia, and I. B. Spielman, Nature (London) {\bf 471}, 83 (2011).  Y.-J. Lin, R. L. Compton, K. Jimenez-Garcia, W. D. Phillips, J. V. Porto, and I. B. Spielman, Nat. Phys. {\bf 7}, 531 (2011); 

\bibitem{2b}A. V. Chaplik and L. I. Magarill, Phys. Rev. Lett. {\bf 96}, 126402
(2006).

\bibitem{YuZhai} Z.-Q. Yu and H. Zhai, Phys. Rev. Lett. {\bf 107}, 195305 (2011).

\bibitem{Iskin} M. Iskin and A. L. Subasi, Phys. Rev. Lett. 107, 050402 (2011).

\bibitem{Shenoy} J. P. Vyasanakere and V. B. Shenoy Phys. Rev. B 83, 094515 (2011)

\bibitem{Subasi} M. Iskin and A. L. Subasi, Phys. Rev. A 84, 043621 (2011).

\bibitem{VictorGalitski} S. Takei, C.-H. Lin, B. M. Anderson, and V. Galitski, Phys. Rev. A 85, 023626 (2012).

\bibitem{lei}L. Jiang, X. -J. Liu, H. Hu, and H. Pu, Phys. Rev. A {\bf 84}, 063618 (2011).

\bibitem{Zhai2010} C. Wang, C. Gao, C.-M. Jian, and H. Zhai, Phys. Rev. Lett. {\bf 105}, 160403 (2010).

\bibitem{vj2011}J. P. Vyasanakere, S. Zhang, and V. B. Shenoy, Phys. Rev. B {\bf 84}, 014512 (2011).

\bibitem{GongZhang} M. Gong, S. Tewari, and C. Zhang, Phys. Rev. Lett. {\bf 107}, 195303 (2011).

\bibitem{HuiPu} H. Hu, L. Jiang, X.-J. Liu, and H. Pu, Phys. Rev. Lett. {\bf 107}, 195304 (2011). 

\bibitem{Zhang} P. Wang, Z.-Q. Yu, Z. Fu, J. Miao, L. Huang, S. Chai, H. Zhai, and J. Zhang, Phys. Rev. Lett. {\bf 109}, 095301 (2012).

\bibitem{mit} L. W. Cheuk, A. T. Sommer, Z. Hadzibabic, T. Yefsah, W. S. Bakr, and M. W. Zwierlein, Phys. Rev. Lett. {\bf 109}, 095302 (2012).

\bibitem{Jacob2007} A. Jacob, P. Ohberg, G. Juzeliunas, and L. Santos, Appl. Phys. B. {\bf 89}, 439, (2007).

\bibitem{RMP} J. Dalibard, F. Gerbier, G. Juzeliunas, and P. Ohberg, Rev. Mod. Phys. {\bf 83}, 1523 (2011).

\bibitem{3DSOC}B. M. Anderson, G. Juzeli\={u}nas, V. M. Galitski, and I. B. Spielman, Phys. Rev. Lett. {\bf 108}, 235301 (2012).

\bibitem{zhou}Y. Li, X. Zhou, and C. Wu, Phys. Rev. B {\bf 85}, 125122 (2012).

\bibitem{deviatedipole} J.-Y. Zhang, S.-C. Ji, Z. Chen, L. Zhang, Z.-D. Du, B. Yan, G.-S. Pan, B. Zhao, Y.-J. Deng, H. Zhai, S. Chen, and J.-W. Pan, Phys. Rev. Lett. {\bf 109}, 115301 (2012). 

\bibitem{ram} B. Ramachandhran, B. Opanchuk, X.-J. Liu, H. Pu, P. D. Drummond, and H. Hu, Phys. Rev. A {\bf 85}, 023606 (2012).

\bibitem{brokeGal}Q. Zhu, C. Zhang, and B. Wu, Europhys. Lett. {\bf 100}, 50003 (2012).

\bibitem{jason} T.-L. Ho and S. Zhang, Phys. Rev. Lett. {\bf 107}, 150403 (2011).

\bibitem{zhaihui} C. Wang, C. Gao, C.-M. Jian, and H. Zhai, Phys. Rev. Lett. {\bf 105}, 160403 (2010). 

\bibitem{yip} S.-K. Yip, Phys. Rev. A {\bf 83}, 043616 (2011).


\bibitem{guo} Z. Zheng, M. Gong, X. Zou, C. Zhang, and G. Guo, Phys. Rev. A {\bf 87}, 031602(R) (2013).

\bibitem{vjnote}V. B. Shenoy, e-print arXiv:1211.1831.

\bibitem{wei} F. Wu, G.-C. Guo, W. Zhang, and W. Yi, Phys. Rev. Lett. {\bf 110}, 110401 (2013).

\end{thebibliography}
\end{document}